# In silico evidence of the relationship between miRNAs and siRNAs

Ludovica Montanucci[1], Piero Fariselli[1*], Pier Luigi Martelli[1], Ivan Rossi[1,2] and Rita Casadio[1]

[1] Biocomputing Group, Department of Biology, via Irnerio 42, University of Bologna, Italy

[2] BioDec Srl, Via Calzavecchio 20/2, I-40033 Casalecchio di Reno (BO), Italy

## ABSTRACT

Both short interfering RNAs (siRNAs) and microRNAs (miRNAs) mediate the repression of specific sequences of mRNA through the RNA interference pathway. In the last years several experiments have supported the hypothesis that siRNAs and miRNAs may be functionally interchangeable, at least in cultured cells. In this work we verify that this hypothesis is also supported by a computational evidence. We show that a method specifically trained to predict the activity of the exogenous siRNAs assigns a high silencing level to experimentally determined human miRNAs. This result not only supports the idea of siRNAs and miRNAs equivalence but indicates that it is possible to use computational tools developed using synthetic small interference RNAs to investigate endogenous miRNAs.

## 1    INTRODUCTION

RNA interference (RNAi), first discovered in the nematode *Caenorhabditis elegans* (Fire *et al.,* 1998), is an evolutionarily conserved endogenous pathway that mediates negative post-transcriptional gene regulation in eukaryotic cells (Hannon 2002; Hannon and Rossi, 2004; Mello and Conte, 2004). The sequence-specific mediators of this regulatory pathway are small (~23 nt) RNA molecules, that bind complementary regions of messenger transcripts and induce highly specific gene silencing (Bartel 2004).

Small RNAs belong to two major classes: microRNAs (miRNAs) and short interfering RNAs (siRNAs). miRNAs originate from short hairpins (called microRNA precursors) endogenously produced through a multi-step process that starts in the nucleus (Carmell and Hannon, 2004). On the contrary siRNAs can be either endogenous or exogenous and derive from double stranded RNA molecules (dsRNAs). miRNAs and siRNAs seem to differ only in their biogenesis (Ambros *et al.,* 2003, Elbashir *et al.*, 2001; Zeng *et al.,* 2003; Doench *et al.,* 2003; Rana 2007).

Long dsRNAs and the miRNA hairpin-shaped precursors are initially processed by the cytoplasmic ribonuclease III-like enzyme Dicer which cleaves them producing dsRNAs about 22 bp long with characteristic 2nt overhangs at the 3' terminus. Then they are incorporated, as single stranded RNA, into a ribonucleoprotein complex named RNA-induced silencing complex (RISC). RISC identifies target genes through an antisense complementarity between the mi/siRNA and the mRNA. When RISC binds the target message, mRNA can be inhibited and later reprocessed or degraded, depending on the degree of pair-matching (Valencia-Sanchez *et al.,* 2006; Rana 2007).

Despite much knowledge has been achieved in the last few years, many aspects of the silencing mechanism are still not fully understood, for instance the perfect theoretical matching complementarity between target message and mi/siRNA does not guarantee the mRNA silencing. On the contrary, also partial matches, as happens in animals (Rhoades *et al.,* 2002; Llave *et al.,* 2002; Tang *et al.,* 2003, Rehmsmeier *et al.* 2004), can lead to processes of gene down-regulation.

As for miRNAs, different siRNAs with perfect base complementarity have different knock-down efficacies, and it is not completely known what determines the highly effectiveness and specificity of some and the weak functionality of others (Reynolds *et al.,* 2004). To this purpose many predictors based on different features have been developed. Among them those based on sequence (Sætrom and Snøve, 2004), thermodynamic properties (Khvorova *et al.,* 2003; Chalk *et al.,* 2004), mRNA secondary structure properties (Luo *et al.,* 2004; Yiu *et al.,* 2005), compositional biases (Henschel *et al.,* 2004), position specific properties (Reynolds *et al.,* 2003; Hsieh *et al.,* 2004). These methods however suffer from poor concordance (Gong *et al.,* 2006) and further studies are needed.

Recently, Huesken and coworkers (Huesken *et al.,* 2005) made available a conspicuous experimental data set on which they implemented a powerful predictor of siRNA activity. Using these data we show that a method specifically trained to predict the activity of synthetic siRNA, assigns a high silencing level to experimentally determined human miRNAs.

## 2    METHODS

### 2.1    Databases

The data set of siRNAs is taken from Novartis (Huesken *et al.,* 2005) and consists of 2431 siRNA sequences whose inhibitory activities had been experimentally determined through a high-throughput reporter assay. We compute our results using the training/testing division indicated in Huesken *et al.* 2005 (the same 2182 siRNA sequences were used for the training and the same 249 siRNAs were used as a test set). We also tested more elaborate (five/ten-folds) cross-validation procedures, but we did not notice any significant difference with respect to the proposed training-testing set.

Furthermore, two databases of miRNAs were selected to investigate the siRNA/miRNA similarity. The first one consists of miRNAs extracted from TarBase that is a database of experimentally supported animal microRNA targets (Sethupathy *et al.,* 2006). From TarBase version 3 (TarBase_V3), we kept the human miRNAs for which the sequences are reported. This set is composed of 36 different miRNAs.

A second dataset was derived from miRBase release 9.0 (Griffiths-Jones *et al.,* 2006). From miRBase we have filtered all human mature microRNAs whose evidence was reported as "experimental". (Mature sequences whose length was less than 21 nucleotides were removed from the set.). The filtered test set contains 349 microRNAs.

## 2.2 The predictor

Similarly to Huesken *et al.* 2005, we used a feed-forward neural network with standard error back propagation training algorithm. In order to improve the accuracy we included into the input window the encoding for: the 21 possible base-positions (like Huesken *et al.,* 2005), the possible nearest neighbor pairs and triplets as proposed before (Vert *et al.,* 2006). This leads to a final neural network input that consists of 164 neurons (the possible base in each 21 positions + the possible 4*4 neighboring base pairs and the 4*4*4 possible neighboring triplets). The network has one output node that codes for the inhibition activity and 8 hidden neurons.

The evaluation of the performance of the prediction method were carried using the Pearson correlation coefficient, defined as:

$$r = \frac{\sum (x_i - \overline{x})(y_i - \overline{y})}{\left[\sum (x_i - \overline{x})^2 \sum (y_i - \overline{y})^2\right]^{1/2}} \tag{1}$$

## 3      RESULTS

When the neural networks was trained using the same input encoding described by Huesken *et al.* 2005, the Pearson correlation coefficient for the test set is exactly the same reported in the original paper, namely 0.66. A slightly improved value is obtained when the input encoding is enriched by the pair and triplet frequencies. In this case we achieved 0.68 (Fig. 1).

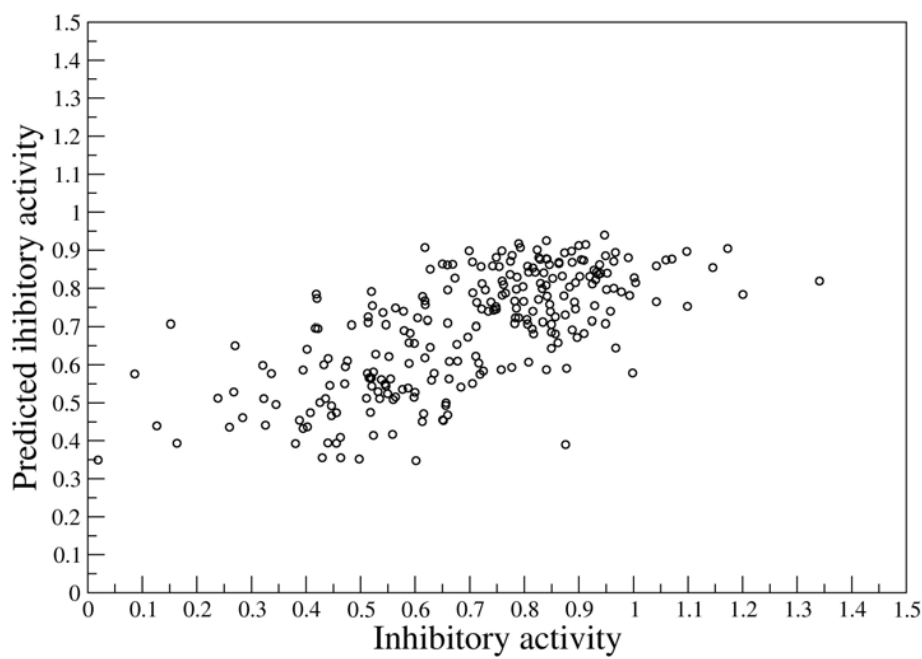

**Fig. 1.** Correlation between observed and predicted inhibitory activities for the test set of 249 siRNAs. The Pearson correlation value is 0.68.

Using this predictor we were in the position of testing *in silico* the hypothesis of chemical-functional equivalence between siRNAs and endogenous miRNAs. In particular, even if the computational method was trained to identify active siRNAs, it assigns high scores also to the endogenous miRNAs. However, the two types of small RNAs maintain a subtle difference: the designed siRNAs have a perfect base-matching with the corresponding mRNAs, while the miRNAs extracted from the data have only a partial base pairing.

Nonetheless, the results reported in Figure 2, confirm the hypothesis, since the neural-network predicts high inhibition activities for most of the experimental miRNAs. This is true for both human microRNAs derived from TarBase and miRBase databases (Fig. 2). In Fig. 2, where we report the distribution of the predicted efficacies for miRNAs, it is evident that the vast majority of miRNAs are predicted with an inhibitory level higher than 0.7.

This findings not only support the idea of siRNAs and miRNAs equivalence but indicates that it is possible to adopt computational tools developed using synthetic small interference RNAs to investigate endogenous miRNAs.

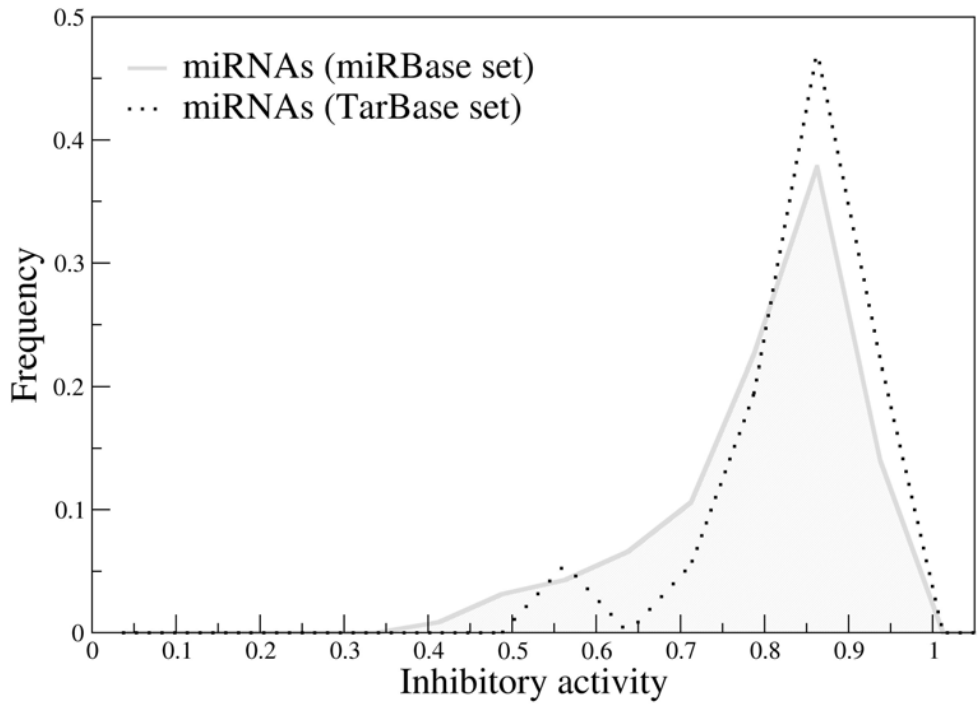

**Fig. 2.** Distribution of predicted inhibitory activities for the two data sets of miRNAs. (from TarBase: dotted line; from miRBase: gray line).

## 4    ACKNOWLEDGEMENTS

This work was supported by the following grants: LIBI-Laboratorio Internazionale di BioInformatica, delivered to RC. IR acknowledges the EU FP6 Specific Targeted Research Project TargetHerpes (LSHG-CT-2006-037517) for support.